\documentclass{emulateapj}
\usepackage{ulem}
\usepackage{longtable}
\usepackage{graphicx}
\input{epsf.sty}
\input{psfig.sty}
\usepackage{url,color}

\shorttitle{IGR J12580+0134: A TDE with Off-beam Relativistic Jet}
\shortauthors{Lei et al.}
\slugcomment{}

\begin{document}
\title{IGR J12580+0134: The First Tidal Disruption Event with an Off-beam Relativistic Jet}
\author{Wei-Hua Lei$^{1,2}$, Qiang Yuan$^3$, Bing Zhang$^2$ and 
Q. Daniel Wang$^3$}
\affil{$^{1}$School of Physics, Huazhong University of Science and Technology, Wuhan 430074, China. Email: leiwh@hust.edu.cn \\
$^{2}$Department of Physics and Astronomy, University of Nevada Las Vegas, NV 89154, USA  \\
$^{3}$Department of Astronomy, University of Massachusetts Amherst, MA 01003, USA. Email: yuanq@umass.edu
}

\begin{abstract}
Supermassive black holes (SMBHs) can capture and tidally disrupt stars 
or sub-stellar objects orbiting nearby. The detections of Sw J1644+57-like 
events suggest that at least some TDEs can launch a relativistic jet 
beaming towards Earth. A natural expectation would be the existence of 
TDEs with a relativistic jet beaming away from Earth. The nearby TDE 
candidate IGR J12580+0134 provides new insights into the jet phenomenon. Combining 
several constraints, we find that the event invokes a $8-40$ 
Jupiter mass object tidally disrupted by a $3 \times 10^5 - 1.8 \times 
10^7 M_\sun$ SMBH. Recently, a bright radio transient was discovered by
Irwin et al. in association with IGR J12580+0134. We perform detailed 
modeling of the event based on a numerical jet model previously developed 
for the radio emission of Sw J1644+57. We find that the radio data of 
IGR J12580+0134 can be interpreted within an external forward shock model 
in the Newtonian regime. Using Sw J1644+57 as a template and properly 
correcting for its luminosity, we argue that the observed X-ray flux in 
early times is too faint to allow an on-beam relativistic jet unless the 
Lorentz factor is very small. Rather, the X-ray emission is likely from 
the disk or corona near the black hole. From various constraints, we find 
that the data are consistent with an off-beam relativistic jet with a 
viewing angle $\theta_{\rm obs} \gtrsim 30^{\rm o}$, and an initial 
Lorentz factor $\Gamma_j \gtrsim $ a few.This scenario can readily be tested in the upcoming VLBI observations.
\end{abstract}

\section{Introduction}
A star or sub-stellar object may be disrupted by tidal forces when 
it passes close enough by a supermassive black hole (SMBH). These events 
--- known as TDEs --- are expected to occur every $10^3 - 10^5$ years for 
a typical galaxy (Magorrian \& Tremaine 1999; Wang \& Merritt 2004). 
The debris of the stellar object will be accreted onto the black hole 
(BH), producing flaring emission in X-ray, ultraviolet, and optical. 
A typical $t^{-5/3}$ behavior of the observed luminosity tracking the
fallback rate evolution of the stellar debris, is a distinctive 
feature of TDEs (Rees 1988; Evans \& Kochanek 1989; Phinney 1989).

The detection of Sw J1644+57 at $z = 0.35$ suggested that at least some 
TDEs can launch a relativistic jet towards Earth, which is manifested 
as a super-Eddington X-ray burst (Bloom et al. 2011; Burrows et al. 2011; 
Levan et al. 2011; Zauderer et al. 2011) and a long lasting radio emission 
resulting from jet-medium interaction (Zauderer et al. 2011; 
Wang et al. 2014; Tchekhovskoy et al. 2014; Liu et al. 2015). Sw J2058+05 
(Cenko et al. 2012) and Sw J1112.2-8238  (Brown et al. 2015) are two 
other candidates that belong to such a category. 
A direct expectation is that there should be TDE relativistic jets that 
beam away from Earth\footnote{Such jets are called off-beam jets, with 
$\theta_{\rm obs} > \max(\theta_j, 1/\Gamma_j)$, where $\theta_{\rm obs}$ 
is the angle between the jet axis and the line-of-sight, $\theta_j$ and
$\Gamma_j$ are the opening angle and the initial Lorentz factor of the 
jet. In contrast, for an on-beam jet, $\theta_{\rm obs} \leq {\rm max}(\theta_j,
1/\Gamma_j)$ is satisfied.}. 

IGR J12580+0134 in the nucleus of NGC 4845 --- a galaxy located at the 
distance of only $\simeq 17 \rm Mpc $ --- is likely such a case. 
IGR J12580+0134 is a flaring hard X-ray source initially observed by 
\textit{Integral} (Walter et al. 2011). Follow-up X-ray observations 
with \textit{XMM-Newton}, \textit{Swift}, and \textit{MAXI}, together 
with the \textit{Integral} data suggested that the source resulted 
from a TDE of a super-Jupiter by the central SMBH in NGC 4845 
(Nikolajuk \& Walter 2013). The X-ray lightcurve after the peak is 
consistent with the $t^{-5/3}$ decay law, as expected by the simple 
TDE picture. The last point at $t\sim 600$ days drops significantly below
the extension of such a power-law. The X-ray lightcurve is similar
to that of Sw J1644+57 (Burrows et al. 2011; Zauderer et al. 2013), but with
less variability observed. The hard X-ray emission was suggested to 
come from a corona forming around the accretion flow 
close to the BH (Nikolajuk \& Walter 2013). The drastic decay
at late time is attributed to a significant drop of the accretion rate,
 or more specifically, when the inner disk changes from the advective 
state to the gas-pressure-dominated state (Shen \& Matzner 2014). 
The fast decline of the X-ray emission suggests that it has a central 
engine origin instead of the shocks due to the jet-medium interaction which 
predicts a much shallower decay (Zauderer et al. 2013; Wang et al. 2014). 

The radio counterpart of the TDE (about one year after its X-ray peak) 
was detected serendipitously by Karl G. Jansky Very Large Array (JVLA) 
in a nearby galaxy survey (Irwin et al. 2015). Compared with the radio 
flux inferred from the VLA observations taken years ago, the central compact 
source was brightened by more than a factor of $\sim 10$. The radio 
spectral shape, peaking at GHz frequencies, and its variation suggest 
self-absorbed synchrotron emission with decreasing optical thickness 
(Irwin et al. 2015). The detection of the $(2-3)\%$ circular polarization 
and no significant linear polarization further supports this scenario 
(Beckert \& Falcke 2002; O'Sullivan et al. 2013). The observational 
properties can be naturally explained by an expanding radio lobe, powered 
by interaction between a jet and a circum-nuclear medium (CNM) of the TDE. 
However, the sub-Eddington feature of the X-ray emission makes it different 
from Sw J1644+57-like on-beam events. An attractive possibility is that 
this is the first off-beam jetted TDE.

An analytical model of the jet-CNM interaction to explain the radio 
emission characteristics has been proposed in Irwin et al. (2015). The model is 
based primarily on order-of-magnitude estimates, especially for the interpretation of X-ray 
emission with the inverse Compton (IC) counterpart of the synchrotron jet. In this work, we apply a relativistic jet model 
to study the radio and X-ray data of IGR J12580+0134, by self-consistently 
modeling the dynamical evolution of the jet and the synchrotron radiation 
properties of the electrons.
The model has been successfully applied to the radio emission of 
Sw J1644+57 (Wang et al. 2014). While the radio emission of IGR 
J12580+0134 is found to be consistent with the external shock synchrotron 
emission in the Newtonian regime, and the early X-ray emission could be from 
1) the disk/corona, 2) the internal dissipation within the jet, and 3) the 
external shock. To satisfy the X-ray constraints, the on-beam internal 
jet dissipation and the external shock X-ray synchrotron emission
need to be strongly suppressed, which implies that the TDE has an off-beam 
jet. However, the shape of the X-ray lightcurve disfavors the external shock scenario
due to the jet-CNM interaction. The internal dissipation may lead to strong variability 
as Sw J1644+57, which is not found in IGR J12580+0134. Its X-ray luminosity and temporal behavior resemble those of typical non-jetted ROSAT TDEs discovered in NGC 5905.
We therefore expect that the early X-ray emission 
is more likely of a disk/corona origin.

The paper is organized as follows. We constrain the masses of the SMBH 
and the disrupted object using the X-ray data in Section 2. The on-beam jet
model and its difficulties are discussed in Section 3. In Section 4, we model
the radio data in detail within the off-beam jet model, and derive the 
constraints on model parameters. The results are summarized in Section 5
with some discussion.

\section{Tidal Disruption of a Jupiter-like Object by a Supermassive Black Hole}

IGR J12580+0134 was discovered by \textit{Integral} (Walter et al. 2011) 
during January 2-11, 2011, with a position consistent with that of a nearby 
spiral galaxy NGC 4845. \textit{Swift}/XRT and \textit{XMM-Newton} 
observations started a few days later, confirmed the association of the transient with 
the nucleus of the galaxy. The peak flux of the transient is $F_{2-10\,
\rm keV} > 5.0 \times 10^{-11}\,\rm erg \,cm^{-2}\, s^{-1}$ 
(Nikolajuk \& Walter 2013), 
corresponding to a brightening by a factor $> 100$ compared with the 
flux upper limit of the galaxy before the outburst (Fabbiano et al.
1992). The sharp onset and the subsequent power law decline with a slope consistent
with $-5/3$ suggest that the transient may be triggered by tidal disruption 
of a star or sub-stellar object by the SMBH (Nikolajuk \& Walter 2013).

Disruption of a star occurs when it comes to a 
BH closer than the tidal disruption radius $R_{\rm{T}}$, which is 
determined through equating the mean density of the volume enclosed by 
$R_{\rm{T}}$ and the density of the star. The tidal disruption radius 
is then given by
\begin{eqnarray}
R_{\rm{T}} \simeq \left(\frac{M_{\bullet}}{M_*}\right)^{1/3} R_* 
\simeq  7 \times 10^{12}\,m_*^{-1/3} r_* M_{\bullet,6}^{1/3}\ {\rm cm}, 
\end{eqnarray}
where $M_{\bullet}$ is the mass of the BH, $M_*$ and $R_*$ are
the mass and radius of the star, respectively, $m_*=M_*/M_{\sun}$, 
$r_* = R_*/R_{\sun}$ are normalized to solar values, and the black mass  
$M_{\bullet,6}=M_{\bullet}/10^6$ M$_{\sun}$ is normalized to million
solar masses.

After the disruption, part of the stellar object is unbound. The bound 
part can lead to flaring electromagnetic emission when it is accreted 
onto the BH after making one more orbit back to pericenter. The time scale
for the first main stream of disrupted materials (those with the lowest energy) 
to return to the pericenter is
\begin{eqnarray}
\Delta t_{\rm m} \simeq \frac{\pi}{2^{1/2}} \left(R_{\rm P}/R_*\right)^{3/2} 
\left(\frac{R_{\rm P}^3}{GM_\bullet}\right)^{1/2}  \nonumber \\ 
\simeq 3.5\times 10^6\,{\rm s} \ M_{\bullet,6}^{1/2} b^{-3} m_*^{-1} r_*^{3/2},
\label{eq:deltat}
\end{eqnarray}
where the impact parameter $b \equiv R_{\rm T}/R_{\rm P}$, defined as the
ratio of the tidal radius $R_{\rm T}$ to the pericenter radius $R_{\rm P}$, 
describes the effective depth of the encounter.

Assuming a ``flat'' mass distribution after the disruption, the rate at which
materials with progressively higher energy to return to their respective 
orbital periastrons after one orbit is
\begin{equation}
\dot{M} = \frac{1}{3} \frac{\Delta M}{\Delta t_{\rm m}} \left(\frac{t-t_{\rm D}}{\Delta t_{\rm m}} \right)^{-5/3}.
\label{eq:dotM}
\end{equation}
This defines the well-known ``fallback'' -5/3 law. 
Here $t_{\rm D}$ is the starting time of the tidal disruption,
$\Delta M= f M_*$ is the mass that falls back to pericenter, which is a
fraction $f$ of the original mass of the disrupted object. By fitting 
the X-ray lightcurve of IGR J12580+0134 with this $-5/3$ power-law form, 
Nikkolajuk \& Walter (2013) found that $t_{\rm D}$ was around October 24, 2010. 
The peak luminosity of the X-ray emission occurred on January 22, 2011, 
suggesting $\Delta t_{\rm m} \simeq 60-100$ days. The mass fraction $f$
was assumed to be 0.5 in Rees (1988), which means half of the debris
is bound. However, recent numerical simulations suggested a smaller 
fraction $f\simeq 0.1$ (Ayal et al. 2000). In the following calculation 
we adopt $f=0.1$, and the impact parameter $b$ is adopted as unity.

\subsection{Mass of the SMBH}

We now estimate the mass of the SMBH in various ways. 
The \textit{XMM-Newton} X-ray ($2-10$ keV) lightcurve indicates a variability 
timescale $\delta t_{\rm min} < 90$ s. Assuming this variability time
scale is defined by the innermost stable circular orbit (ISCO) of the
accretion disk, $R_{\rm in}/c$, the BH mass can be estimated as
\begin{equation}
M_{\bullet,6} \simeq 18\, r_{\rm in}^{-1}
\left(\frac{\delta t_{\rm min}}{90\, \rm s}\right)~.
\end{equation}
where, $r_{\rm in} \equiv R_{\rm in}/R_{\rm g}$ is the radius of 
the ISCO in terms of $R_{\rm g}=GM_{\bullet}/c^2$, and can be expressed
as (Bardeen et al. 1972),
\begin{eqnarray}
r_{\rm in} =  3+Z_2 -\left[(3-Z_1)(3+Z_1+2Z_2)\right]^{1/2},
\end{eqnarray}
where $Z_1 \equiv 1+(1-a_{\bullet}^2)^{1/3} [(1+a_{\bullet})^{1/3}+(
1-a_{\bullet})^{1/3}]$, $Z_2\equiv (3a_{\bullet}^2+Z_1^2)^{1/2}$, 
and $a_{\bullet}=J_\bullet c/(G M_{\bullet}^2) \in [0,1]$ is the spin 
parameter of the BH. We then find that $M_{\bullet,6} \lesssim 18$ for 
$a_{\bullet} \lesssim 1$. 

The \textit{XMM-Newton} data revealed a quasi-periodic oscillation (QPO) with frequency 
$\sim 10^{-3}$ Hz (Nikolajuk \& Walter 2013). Assuming that the QPO corresponds 
to the Kepler rotation at the ISCO, we have
\begin{equation}
\nu_{\rm QPO} = \frac{\Omega_D}{2\pi} = \frac{c^3}{2\pi G M_\bullet} \frac{1}{r_{\rm in}^{3/2}+a_\bullet},
\end{equation}
i.e, 
\begin{equation}
M_{\bullet,6} \simeq 32 \frac{10^{-3}\,\rm Hz}{\nu_{\rm QPO}} \frac{1}{r_{\rm in}^{3/2}+a_\bullet}.
\end{equation}
One then has $M_{\bullet,6} \simeq 16$ for $a_{\bullet} \lesssim 1$. This sets an upper limit on the BH mass.

There is also an empirical relation between such a QPO frequency and
the SMBH mass in active galactic nuclei (AGN; Remillard \& McClintock 2006; 
Bian \& Huang 2010):
\begin{equation}
\nu_{\rm QPO} =  \frac{0.931\times 10^{-3}}{M_{\bullet,6}}.
\end{equation}
Applying this relation to IGR J12580+0134, we obtain $M_{\bullet,6} 
\sim 1$.

An independent constraint on the BH mass can be obtained from 
the relation between the B-band luminosity of the bulge and the SMBH 
mass (Kormendy \& Gebhardt 2001; H{\"a}ring \& Rix 2004). The bulge
luminosity of NGC 4845 (Ho, Filippenko \& Sargent 1997) results in a BH 
mass of $13-20\times10^6 M_\sun$. Another empirical relation among the 
radio luminosity at 5 GHz, X-ray luminosity in the 2-10 keV band, and 
the SMBH mass $\log L(5~{\rm GHz}) = (0.82 \pm 0.08) \log M_\bullet + 
(0.62 \pm 0.10) \log L(2-10~{\rm keV}) + (6.31 \pm 0.21)$ 
(M{\"i}ller \& G{\"u}ltekin 2011) gives a rough estimate of the SMBH 
mass of $\sim 10^6 M_\sun$.
Using the relationship between the BH mass and the normalized X-ray 
excess variance, Nikolajuk \& Walter (2013) found a mass of $\sim 3\times 
10^5$ M$_\sun$. 
Using the the relation between the luminosity and width of 
$H_\alpha$ emission (Greene \& Ho 2005; Ho, Filippenko \& Sargent 1997),  
we find a lower limit of the SMBH mass of $3\times 10^5$ M$_\sun$. 
Considering all these constraints, we expect the BH mass to be in the 
range of $3\times 10^5 - 18 \times 10^6 M_\sun$, which is shown in 
Fig. \ref{fig:mass}.

\subsection{Mass of the disrupted object}

The peak time ($\sim \Delta t_{\rm m}$), together with the peak luminosity 
of X-ray emission, can be used to constrain the mass of the disrupted 
object ($M_*$). The peak X-ray luminosity allows us to estimate the peak 
accretion rate. We first assume that the X-ray emission comes from the 
accretion disk and its associated corona. A possible jet origin will 
be discussed in Sections 3 and 4, which is found to be difficult to explain
the data.

For a thin disk, the total (thermal) luminosity from the disk is given by
\begin{equation}
L_{\rm disk} = \epsilon \dot{M}c^2 = (1-E_{\rm in}) \dot{M} c^2
\end{equation}
where $\epsilon$ is the efficiency, and $E_{\rm in}$ is the specific 
energy corresponding to the inner edge radius $r_{\rm in}$. The expression 
for $E_{\rm in}$ is (Novikov \& Thorne 1973; Wang et al. 1998)
\begin{equation}
E_{\rm in} = \frac{4\sqrt{ r_{\rm in} }-3a_{\bullet}}{\sqrt{3} r_{\rm in}},
\end{equation}
For $0< a_\bullet <1$, we have $0.06< \epsilon <0.42$.

The peak of the observed $17.3-80$ keV luminosity is $L_{\rm X,iso}^{\rm peak} 
\simeq 1.5 \times 10^{42} \rm erg \ s^{-1}$, which may be of a non-thermal
corona origin (Nikolajuk \& Walter 2013). 
Assuming that the thermal emission is ten times brighter than this hard 
X-ray emission, the maximum tidal flare luminosity is then 
$L_{\rm flare}/L_{\rm Edd} \simeq 0.1$ for a $10^6$ M$_\sun$ SMBH. 
The peak accretion rate can be estimated as
\begin{equation}
\dot{M}_{\rm peak} = \frac{10\,L_{\rm X,iso}^{\rm peak}}{(1-E_{\rm in}) c^2}  =8.4 \times 10^{-12} (1-E_{\rm in})^{-1} M_\sun \ s^{-1}.
\label{eq:dMp}
\end{equation}
For $0 \leq a_{\bullet} \leq 1$, we have $8.4 \times 10^{-12} < 
\dot{m}_{\rm peak} \equiv \dot{M}_{\rm peak}/{\rm M_{\odot}\,s^{-1}} < 1.5 
\times 10^{-10}$. From equation (\ref{eq:dotM}), we have $\dot{M}_{\rm peak} 
= f M_*/(3\Delta t_{\rm m})$.

\begin{figure}[ht]
\centering
\includegraphics[width=80mm]{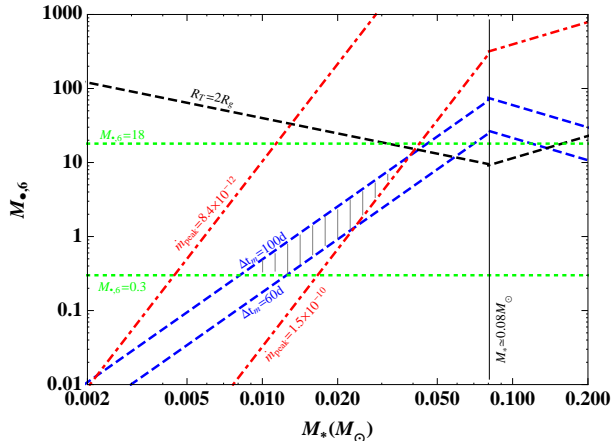} \\
\caption{Constraints on the mass of the BH ($M_\bullet$) and the mass of 
the disrupted object ($M_*$), derived from the X-ray observations of 
IGR J12580+0134. The shaded area shows the allowed region. 
The limits shown include (1) $60\,{\rm d} <\Delta t_{\rm m} <100\,{\rm d}$ 
(blue dashed lines); (2) $8.4 \times 10^{-12} < \dot{m}_{\rm peak} 
< 1.5 \times 10^{-10}$ (red dot-dashed lines); (3) $R_{\rm T} > 2 R_{\rm g}$ 
(black dashed line and below); (4) $0.3 < M_{\bullet,6}<18$ (green 
dotted lines).}
\label{fig:mass}
\end{figure}

Equations (\ref{eq:deltat}), (\ref{eq:dotM}) and (\ref{eq:dMp}) relate the masses of the BH
and the disrupted object to the observables $\Delta t_m$ and 
$\dot{M}_{\rm peak}$. To constrain the mass of the disrupted object, we 
need the mass-radius relation. We consider two possibilities of the 
disrupted object: 1) a low mass star with $R_* = R_\sun (M_*/M_\sun)$ for 
$0.08\,M_\sun < M_* < 1\,M_\sun$, and 2) a substellar object (including 
brown dwarfs and planets) with $R_* \simeq 0.06 R_\sun (M_*/M_\sun)^{-1/8}$ 
for $0.001\,M_\sun < M_* < 0.08\,M_\sun$ (Chabrier \& Baraffe 2000)\footnote{
Chabrier \& Baraffe (2000) stated that their fit is good down to 
$M_* = 0.01 M_\sun$, but a comparison with their plot shows that the fit 
is acceptable even for lower masses. We apply their scaling down to 
$M_* \sim 0.001 M_\sun $. }.
Substituting these mass-radius relations into the Eqs. (\ref{eq:deltat}) 
and (\ref{eq:dotM}), we get blue, dashed boundaries ($\Delta t_m$ constraint)
and red, dash-dotted boundaries ($\dot m_{\rm peak}$ constraint) in Fig. \ref{fig:mass}. 
An additional constraint, namely a lower
limit on the tidal radius $R_{\rm T} > 2 R_{\rm g}$, is also shown in this
figure (Li et al. 2002, black, dashed line). Imposing the range of BH mass 
(green, dotted lines) derived above, we finally get
$M_{\bullet} \sim 3 \times 10^5 - 12 \times 10^6\,M_\sun$, and 
$M_* \sim 0.008 - 0.04\,M_\sun$ or $8-40$ Jupiter mass ($M_{\rm J}$),
as denoted as the shaded area in \ref{fig:mass}. 
Therefore, our results suggest a super-Jupiter (8-40 $M_{\rm J}$) being disrupted 
by a 0.3-18 million $M_\odot$ black hole, which is consistent with the parameters
inferred by Nikolajuk \& Walter (2013).  

\section{X-ray emission from an on-beam jet?}

\begin{figure}[ht]
\centering
\includegraphics[width=60mm]{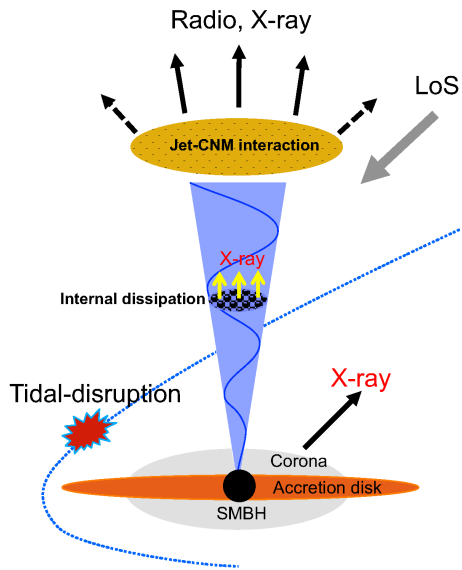} \\
\caption{A schematic illustration of the model, consisting of an accretion 
disk around the SMBH in NGC 4845, and an off-beam relativistic jet. 
``LoS'' denotes the direction of the line-of-sight. X-ray emission can be expected from the
disk, corona, and in the jet (through internal dissipation and jet-CNM interaction). The jet components are 
strongly suppressed at early times due to the large Lorentz factor and view angle.
The observed hard X-ray emission 
comes primarily from an accretion disk and its associated corona near 
the SMBH. Bright radio emission, originated from the external shock due to interaction between 
the jet and the CNM, will be observed when the jet enters the Newtonian phase (emission shown with the dashed arrows). }
\label{fig:model}
\end{figure}

As we have mentioned in the Introduction, the X-ray emission may have
three origins: the disk/corona, internal dissipation, and the jet-CNM
interaction (see Fig. \ref{fig:model} for a cartoon to show the
emission structure of the source). In this section, we discuss the 
possibility of an on-beam jet origin of the X-ray emission. Because the 
X-ray emission is sub-Eddington, one does not have direct information 
about the on-beam jet emission. 
We thus use the {\it scaled} jetted TDE source Sw J1644+57 as a template of the 
emission to develop a constraint\footnote{Although the masses of
the disrupted object and SMBH are very different for these two sources, the jet
physics is likely similar. In the literature, similarities in jet properties between AGNs and
GRBs have been reported (e.g. Wu et al. 2011; 2015). Sw J1644+57 has the best 
coverage of X-ray observations, so is an ideal template for jetted TDEs.}. 
A possible origin of the emission is the internal 
dissipation through, e.g., magnetic reconnection (Zhang \& Yan 2011), 
within the jet.

At $\sim 383$ d, the radio emission flux is $\sim$ 4.37 mJy at 1.8 GHz 
for Sw J1644+57, and is $\sim$ 211 mJy at 1.57 GHz for IGR J12580+0134. 
Considering the distances of the sources, we have the jet luminosity 
$\sim 3.2\times 10^{40} \rm erg\ s^{-1}$ at 1.8 GHz for Sw J1644+57, and
$1.1 \times 10^{38} \rm erg\ s^{-1}$ at 1.57 GHz for IGR J12580+0134. 
Because late time radio emission is a good indicator of the total energy of 
the jet and the time scales of the two events are comparable, 
this comparison implies that the jet power of IGR J12580+0134 is roughly 
$\sim 300$ times weaker than that of Sw J1644+57. This difference factor is also consistent 
with the estimated masses of the disrupted objects in the two events: 
while the mass of the disrupted object in IGR J12580+0134 is 8-40 
$M_{\rm J}$, that of Sw J1644+57 is of the order of solar mass 
(Burrows et al. 2011; Bloom et al. 2011; Lei \& Zhang 2011; Lei et al. 2013). Therefore, the mass 
ratio of the order $\sim 100$ is reflected in the jet power difference of the two events.

\begin{figure}[ht]
\centering
\includegraphics[width=80mm]{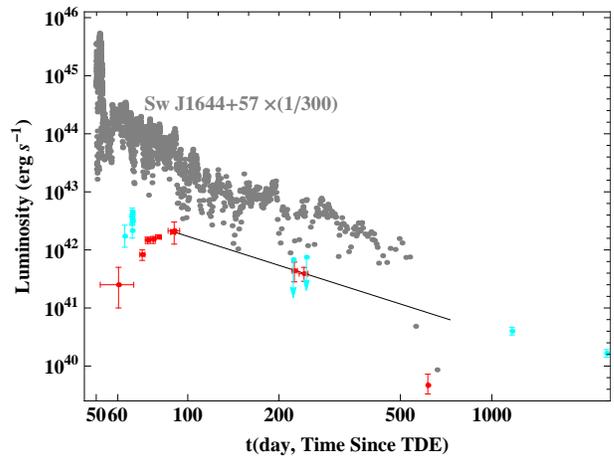} \\
\caption{The X-ray emission in IGR J12580+0134 (red) compared with the X-ray 
data of Sw J1644+57 scaled down by a factor of 300 (gray). The observational 
data  of IGR J12580+0134 (17.3-80 keV) are adopted from Nikolajuk \& Walter 
(2013). The X-ray lightcuve of Sw J1644+57 
is rescaled from the 1-10 keV band to the 17.3-80 keV one for comparison with IGR J12580+0134. The zero point time of IGR J12580+0134 is taken as the time when 
the tidal disruption occurred (2010-10-24 as suggested in Nikolajuk \& 
Walter (2013)), and the zero point time of Sw J1644+57 is taken to be 50 
days before the first \textit{Swift}/BAT trigger. The solid line shows
the $t^{-5/3}$ power law fit. For comparison, we also show a ROSAT 
candidate in NGC 5905 (in 0.1-2 keV)} with cyan (the data were taken from Li et al. 2002), which is a typical non-jetted TDE. 
This suggests that the hard X-ray emission may be dominated by that from disk/corona. 
The contribution from internal dissipation should fall below it due to the strong suppression from an off-beam jet.
\label{fig:onbeam}
\end{figure}

The comparison of the X-ray emissions between Sw J1644+57 (scaled down
by a factor of $1/300$; gray points) and IGR J12580+0134 (red, in energy band 17.3 - 80 keV) is shown in 
Fig. \ref{fig:onbeam}. To compare with the lightcurve of IGR J12580+0134, we rescale the 1 - 10 keV luminosity of Sw J1644+57 to 17.3-80 keV one (by using an average photon index $\sim 1.8$, see Burrows et al. 2011). One can see that the observed emissions are 
substantially lower than what would be expected from an on-beam jet. The peak of the observed 17.3 - 80 keV luminosity is $ \sim 1.5 \times 10^{42} \rm{ erg \ s^{-1}}$ for IGR J12580+0134, and is $ \sim 3.25 \times 10^{48} \rm{ erg \ s^{-1}}$ for Sw J1644+57.
Therefore, the observed peak luminosity of IGR J12580+0134 is $\sim$ 7200 times fainter than what might be
expected for an on-beam jet. Therefore, this scenario is disfavored unless the jet 
power is more than 7200 times lower than an Sw J1644+57 equivalent.

\begin{figure}[ht]
\centering
\includegraphics[width=80mm]{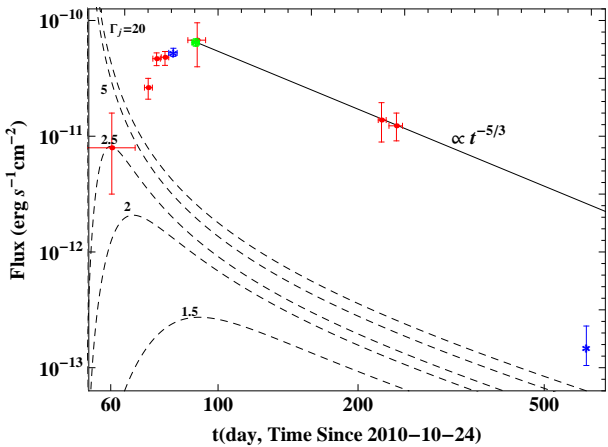}
\caption{The X-ray lightcurve of IGR J12580+0134 compared with the model 
predictions of on-beam jets (dashed lines). The observational 
data from \textit{Integral} (red points), \textit{XMM-Newton} (green star) 
and \textit{Swift}/XRT (blue star) are adopted from Nikolajuk \& Walter 
(2013).
The dashed lines show the lightcurves expected from the external shock in 
the on-beam jet model ($\theta_{\rm obs}=0^{\rm o}$) assuming different 
values of the initial Lorentz factor $\Gamma_j$.}
\label{fig:Xgm}
\end{figure}

Another argument against a relativistic on-beam jet for IGR J12580+0134 
is from the upper limit of the external shock X-ray emission due to the jet-CNM 
interaction. The model lightcurve is expected to reach a peak at the 
deceleration time, and decay with a power law (see Section 4 for a detailed
discussion on the modeling). An on-beam relativistic jet with initial 
Lorentz factor $\Gamma_j$ greater than a few would produce too bright 
X-ray emission to be consistent with the data.
In Fig. \ref{fig:Xgm}, we plot the predicted X-ray lightcurves (see details 
in Section 4 for the parameters to reproduce the radio data, as shown in 
Table 1) for an on-beam jet with different initial Lorentz factors,
and compare them with the observational data.
One can see that  $\Gamma_{\rm j}$ should be less than 2.5 in order not 
to exceed the data. This is another argument against an on-beam 
relativistic jet.

The jet associated with this TDE must be off-beam. As we will show in 
Section 4, the X-ray lightcurve expected from the external shock model 
is inconsistent with the data. Thus the X-ray emission of IGR J12580+0134 
should be either from the disk/corona or the internal dissipation within 
an off-beam jet. In both cases the X-ray luminosities are related to 
the accretion rate of the disrupted debris. The peak time and rising 
slope, which may be related to the orbital periods of the main bound 
material of the disrupted object, of IGR J12580+0134 differ significantly 
from that of Sw J1644+57. This may be due to different orbits of these 
two cases. The X-ray emission of Sw J1644+57 is expected to be from the 
internal dissipation, as manifested by the super-Eddington luminosity 
(Doppler boosted) as well as the significant variabilities. For IGR 
J12580+0134, the X-ray luminosity is sub-Eddington, and the short time 
variabilities are not observed due to the sparse observational data. 
In Fig. \ref{fig:onbeam}, we also show the lighutcurve of a typical non-jetted 
TDE candidate NGC 5905 (cyan). The similarity in their X-ray behavior suggests 
that the hard X-ray emission may be dominated by that from disk/corona. 
However, the internal dissipation is also allowed even though the disk/corona scenario gives 
the most nature explanation.

\section{Off-beam Jet Model}

Radio emission from IGR J12580+0134 has been detected by JVLA at 1.57 and 
6 GHz about one year after the X-ray peak (Irwin et al. 2015). The energy 
spectra, polarization properties, and the time evolution of the radio 
emission suggest a self-absorbed synchrotron emission origin from an 
expanding radio lobe powered by the jet associated with the TDE. In this 
work we study the jet dynamics in detail with a numerical model of the jet 
evolution (Huang et al. 2000) and external shock emission. The model is used
to fit the lightcurves and spectra in radio. As discussed in Section 2, the
X-ray emission is likely of a disk/corona origin.
Nevertheless, we can use the emission as an upper limit of the external shock 
flux to constrain model parameters. 

The model we adopt was developed by Wang et al. (2014), which successfully 
interpreted the late time radio data of Sw J1644+57. We consider a jet 
with opening angle $\theta_{\rm j}$, isotropic kinetic energy $E_{\rm k, iso}$ and initial Lorentz factor $\Gamma_{\rm j}$ 
propagating into a CNM with a constant proton number density $n$. The jet first undergoes a coasting phase, where the jet moves at a nearly constant speed. It starts to decelerate when the mass of the CNM swept by the forward shock is about $1/\Gamma_{\rm j}$ of the rest mass in the ejecta. Then the jet evolves into the second phase. Finally, the blastwave enters the Newtonian phase when it has swept up the CNM with the total rest mass energy comparable to the energy of the ejecta. In this phase, the velocity is much smaller than the speed of light. During all these three phases, electrons are believed to be accelerated
at the forward shock front to a power-law distribution $N(\gamma_{\rm e}) \propto \gamma_{\rm e}^{-p}$. A fraction $\epsilon_{\rm e}$ of the shock energy is distributed into electrons, while another fraction $\epsilon_{\rm B}$ is in the magnetic field generated behind the shock. Accounting for the radiative cooling
and the continuous injection of new accelerated electrons at the shock front, one expects a broken power-law energy spectrum of them, which leads to a multi-segment broken power-law radiation spectrum at any epoch (see Gao et al. 2013 for a detailed review). The evolution features and radiation properties of the jet during the three phases are described in the Appendix. 

In our numerical code, the dynamical evolution of the jet is described by a set of hydrodynamical equations (Huang et al. 2000). The synchrotron spectra of the jet 
are calculated following the standard broken-power-law spectral model
developed for gamma-ray bursts (GRBs; see Gao et al. 2013 for a detailed review). We also used the corrections introduced by Sironi \& Giannios (2013) for the ``deep Newtonian phase'', when the bulk of the shock-accelerated electrons are non-relativistic (see also Huang \& Cheng 2003).  In order to 
give a smooth fit to the radio data, we characterize the spectra around 
the self-absorption frequency $\nu_{\rm a}$ as
\begin{equation}
F_\nu =  F_{\nu}^{\rm thick} (1-e^{-\tau}),
\label{eq:Fva}
\end{equation}
where $F_{\nu}^{\rm thick}$ is the flux at $\nu \ll \nu_{\rm a}$, and
$\tau$ is the optical depth, defined as
\begin{eqnarray}
\tau =\left(\frac{\nu}{\nu_{\rm a}}\right)^{-\alpha}.
\end{eqnarray}
For the slow cooling case, we have $\alpha=5/3$ for $\nu_{\rm a}<\nu_{\rm m}$; $\alpha=(p+4)/2$ 
for $\nu_{\rm m}<\nu_{\rm a}$; and $\alpha=(p+5)/2$ for $\nu_{\rm a}>\nu_{\rm c}$. The critical 
frequencies $\nu_{\rm a}$, $\nu_{\rm m}$ and $\nu_{\rm c}$ in different dynamical regimes
are exhibited in the Appendix.  

We define the time of the first observation of the X-ray outburst (December 
12, 2010) as the starting time ($t_0$) of the jet (the initial disruption 
occurred $\sim 50$ days earlier; Nikolajuk \& Walter 2013). Energy
injection from the central engine follows $t^{-5/3}$ law according to the
X-ray lightcurve. In such a case, the late time dynamics of the jet only 
depends on the total ejected kinetic isotropic-equivalent energy 
$E_\mathrm{k,iso}$, which is about the energy injected in the initial 
emission episode (Zhang \& M\'esz\'aros 2001).

For a collimated jet, the jet break effect becomes important when 
$1/\Gamma > \theta_\mathrm{j}$, where $\Gamma$ is the Lorentz factor and 
$\theta_j$ is the opening angle of the jet (Zhang \& M\'esz\'aros 2004). 
After the jet break, we include a suppression of the flux density by a factor 
of $(\Gamma \theta_\mathrm{j})^2 /2$ (Zhang \& M\'esz\'aros 2004).

The observed flux density is further subject to a correction factor due to 
the viewing angle for an off-beam observer (e.g., Granot et al. 2002)
\begin{equation}
F_{\nu}(\psi,t) = a_{\rm off}^3 F_{\nu /a_{\rm off}}(0,a_{\rm off} t),
\label{eq:Fvoff}
\end{equation}
where $\psi=\rm{max}(0,\theta_{obs}-\theta_j)$ is the angle between the 
near-edge of the jet and the observer, and 
\begin{equation}
a_{\rm off}=\frac{{\mathcal D}_{\rm off}}{{\mathcal D}_{\rm on}}=
\frac{1-\beta}{1-\beta \cos \psi}, 
\label{eq:aoff}
\end{equation}
is the ratio of the off-beam Doppler factor to the on-beam Doppler factor,
with $\beta=\sqrt{1-1/\Gamma^2}$.

\begin{figure}[ht]
\centering
\includegraphics[width=80mm]{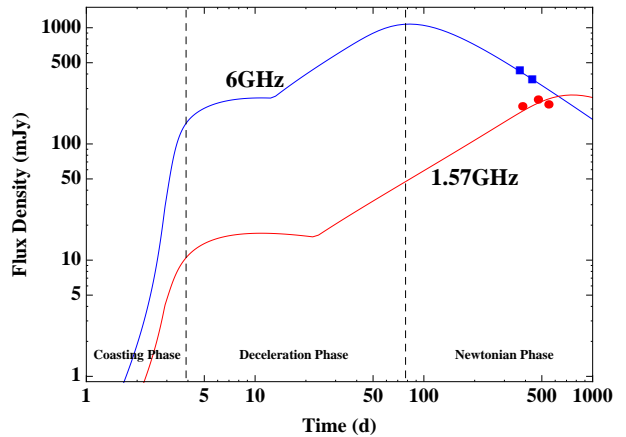} \\
\caption[]{Radio lightcurves of IGR J12580+0134 extending to $t \simeq 600$ 
days at 1.57 GHz (red) and 6 GHz (blue), respectively. The observational 
data are adopted from Irwin et al. (2015). As shown in Appendix, the two 
critical times (the vertical lines), i.e., the deceleration time 
$t_{\rm dec}$ and the Sedov time $t_{\rm Sedov}$, divide the space into 
three phases: the coasting phase, the deceleration phase and the Newtonian 
phase.}
\label{fig:lc}
\end{figure}

\begin{figure}[ht]
\centering
\includegraphics[width=80mm]{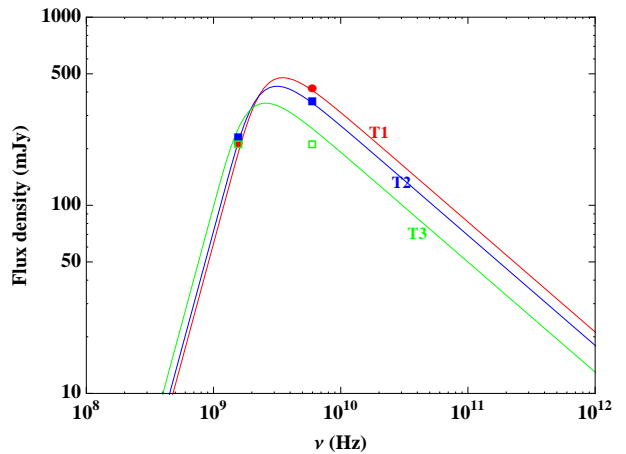} \\
\caption[]{Multi-frequency radio spectral distributions of IGR J12580+0134 
at $t \simeq$ 383 (T1, red), 439 (T2, blue), and 579 (T3, green) days. 
Note that the data (squares) were not obtained simultaneously 
for different bands. They were interpolated or extrapolated to these three
epochs (Irwin et al. 2015). This may account for the slight deviation of the C-band flux at T3 from the model prediction.}
\label{fig:spec}
\end{figure}

The model is characterized by a set of parameters: the density of the
CNM $n$, the observer's view angle $\theta_{\rm obs}$, the isotropic-equivalent 
injected kinetic energy of the jet $E_{\rm k,iso}$, the initial Lorentz factor 
$\Gamma_j$, the jet opening angle $\theta_j$, the spectral index of accelerated
electrons $p$, and the energy density fractions of electrons ($\epsilon_{e}$) 
and the magnetic field ($\epsilon_{\rm B}$).
The jet dynamical evolution equations are solved numerically. An analytical 
description of the main properties of the jet evolution is given in the
Appendix. We apply the model to simultaneously fit the radio data observed 
at three epochs, i.e., 30-Dec-2011 (T1), 24-Feb-2012 (T2) and 13-Jul-2012 
(T3), and two frequencies of 1.57 and 6.0 GHz (Irwin et al. 2015). 
Figs. \ref{fig:lc} and \ref{fig:spec} show an illustration of the model 
expectations compared with the data, for one set of the model parameters 
as presented in Table 1. 

\begin{table*}[htp]
\begin{center}
\caption{The set of parameters adopted to perform the radio data fitting \label{tb:fit}}
\begin{tabular}{cccccccc}
\hline\noalign{\smallskip}
\hline\noalign{\smallskip}
    $n$ (${\rm cm}^{-3}$) & $\theta_{\rm obs}$ (deg) & $E_{50}$ & $\Gamma_{\rm j}$ & $\theta_{\rm j}$ (deg) & $p$ & $\epsilon_{\rm B}$ & $\epsilon_{\rm e}$ \\
\hline\noalign{\smallskip}
3.5 & 34 & 35.0   & 9 & 8.0 & 2.17 & 0.24 & 0.27 \\ 
\noalign{\smallskip}\hline
\end{tabular}
\end{center}
\end{table*}

The electron spectral index $p$ is determined through the in-band spectral
measurements of the radio emission at 6 GHz (C-band), at which the emission is
expected to be optically thin (Irwin et al. 2015). The C-band index 
$\sim -0.587$ at T3 suggests an electron spectral index of $p\simeq 2.17$. 
The low frequency 
(L-band at 1.57 GHz) emission is, however, in the optically thick regime,
resulting in a turn-over of the spectra as shown in Fig. \ref{fig:spec}. 
See Eq. (\ref{eq:Newton_nu}) for the analytical expression of the 
self-absorption frequency of the late stage jet evolution. The peak of the 
radio spectrum declines and shifts to lower frequencies with increasing
time, which can be understood in terms of Eq. (\ref{eq:Newton_nu}). 
Our model can explain the general evolution trend of the L-band and 
C-band spectral indices. 

For other parameters, we have only loose constraints.
The total kinetic energy of the jet should be smaller than the 
mass of the disrupted object, which is $<7\times 10^{52}$ erg. The Lorentz 
factor for relativistic jetted TDEs is expected to be the order of a few 
to a few tens (Metzger et al. 2012; Wang et al. 2014). The energy fractions 
$\epsilon_e$ and $\epsilon_B$ are expected to be $\sim 0.33$ according to 
the equipartition condition. However, GRB afterglow modeling gives a wider
distribution in the relativistic phase (e.g. Kumar \& Zhang 2015 for a review). 
With the radio data alone, one cannot well 
constrain the model parameters, due to the strong degeneracies among the
parameters, especially in the late deep Newtonian stage. 

\begin{figure}[ht]
\centering
\includegraphics[width=80mm]{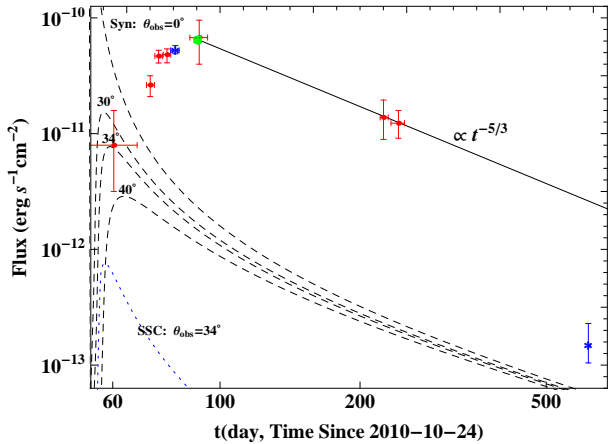}
\caption{The X-ray lightcurve of IGR J12580+0134 compared with model 
predictions of off-beam jets with different view angles (dashed lines). 
The starting time of the jet emission is December 12, 2010, i.e., the 
jet launching time we define. For comparison, we also plot the expected 
SSC emission from the external shock for $\theta_{\rm obs}=34^\circ$ 
(dotted line).}
\label{fig:xray}
\end{figure}

We note that the early X-ray emission can give effective constraints on
the observer's view angle $\theta_{\rm obs}$. Figure \ref{fig:xray} shows 
the expected synchrotron X-ray lightcurves from the external shock of the
jet-CNM interaction, for different $\theta_{\rm obs}$. Other parameters
are the same as those listed in Table 1, which can still reproduce the 
radio data. (In the Newtonian regime, the observer's viewing angle does 
not make a difference in the observed flux.)
We find that $\theta_{\rm obs}$ needs to be large enough in order not to
over-produce the X-ray emission at the early stage. For this 
particular set of parameters we find $\theta_{\rm obs}\gtrsim 30^{\circ}$.

\begin{figure}[ht]
\centering
\includegraphics[width=80mm]{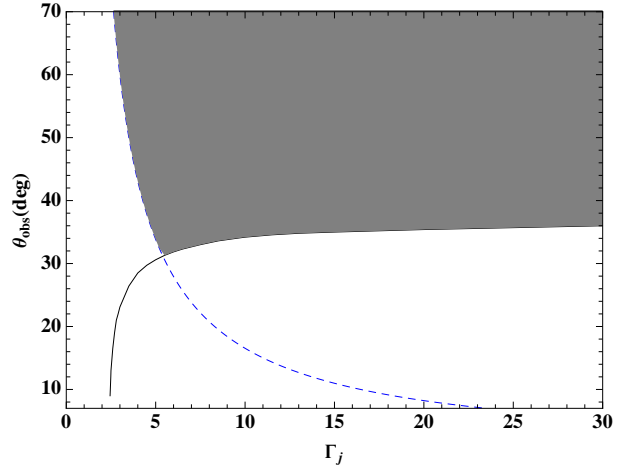}
\caption{Constraints on the $\Gamma_j-\theta_{\rm obs}$ parameter space,
from the X-ray lightcurve as shown in Fig. \ref{fig:xray} (above the solid 
line), and the off-beam factor $a_{\rm off}^4 < 1/7200$ (above the dashed 
line). The shaded region shows the allowed parameter space satisfying both 
constraints. Other parameters are the same as those in Table 1.}
\label{fig:theta}
\end{figure}

Since $\Gamma_j$ is unknown, the above X-ray
constraint only defines a regime in the $\Gamma_j - \theta_{\rm obs}$ parameter
space (region above the black solid curve in Fig. \ref{fig:theta}).
In Section 3, we derived another constraint on $\theta_{\rm obs}$ and 
$\Gamma_{\rm j}$. The peak X-ray luminosity $1.5 \times 10^{42} \rm erg\ 
s^{-1}$ should be at least $\sim 7200$ times lower than the on-beam flux. 
This places an upper limit of the off-beam factor $a_{\rm off}^4$ 
(because we compare $\nu F_{\nu}$ here, see Eq. (\ref{eq:Fvoff})) to be 
$< 1/7200$. This corresponds to the region above the blue dashed curve 
in Fig. \ref{fig:theta}. Combining the two constraints, one gets a shaded 
area in Fig. \ref{fig:theta}, which roughly corresponds to $\Gamma_{\rm j} 
\gtrsim$ a few and $\theta_{\rm obs} \gtrsim 30^\circ$.

With our fitting parameters, the SSC emission is 
less important than the synchrotron emission from the external shock, 
as shown by the dotted line in Fig. \ref{fig:xray}.
The major argument against the SSC explanation is the shape of the X-ray 
lightcurve. As shown in Fig. \ref{fig:xray}, the lightcurve from an off-beam 
jet (similar in shape for both the synchrotron and SSC components) is 
difficult to reproduce the broad peak of the data. An on-beam jet with a very 
small Lorentz factor could in principle match the shape\footnote{Compared
with the on-beam case, the oberved time of the off-beam source will be
diluted by factor $a_{\rm off}^{-1}$. The factor $a_{\rm off}^{-1}$
decrease with time due to the deceleration of the jet, which makes the
lightcurve steeper for off-beam source.}. However, in this case the flux 
is too low when compared with the data. 
The late time X-ray observations with \textit{Swift}/XRT and 
\textit{Integral} revealed a 
drastic decline in the lightcurve evolution at $t>200-600$ days, 
compared with the $t^{-5/3}$ power-law. This behavior is also very 
different from the expectation of the external shock emission which shows 
a much shallower decay. A rapid decline in X-ray emission was also observed 
in Sw J1644+57 (Zauderer et al. 2013), which can be interpreted as the drastic 
decrease of the accretion rate and suggests different sites of the X-ray 
and radio emissions. 
The same argument can be applied to IGR J12580+0134. The potential QPO
observed in X-rays (Nikolajuk \& Walter 2013) provides additional hint 
that the X-ray emission may be related to the accretion disk instead of 
the jet. We note that the last \textit{Swift} point at $t\sim 600$ days 
is not too far away from the external shock model expectation. A deep 
observation by Chandra or NuSTAR may detect the external shock synchrotron 
X-ray component.

\section{Conclusions and Discussion}

The nearby TDE IGR J12580+0134, discovered in X-rays by \textit{Integral} 
and then detected in radio by JVLA, likely launched a relativistic jet. 
In this work, we establish a dynamical jet evolution model to interpret
the radio observations of the source, and derive constraints on the
physical parameters of the TDE with multi-wavelength data. We find
that the BH mass is in the range $M_{\bullet} \sim 3 \times 10^5 - 12 
\times 10^6 M_\sun$, and the mass of the disrupted star is in the range 
$M_* \sim 0.008 - 0.04\,M_\sun$ or $8 -40$ Jupiter mass. The radio 
lightcurves and multiband spectra can be well explained by the synchrotron 
emission from the external shock of the jet, in good agreement with the
conclusion reached by Irwin et al. (2015). The turn-over in the spectra 
is due to the synchrotron self-absorption. The evolution of the peak 
frequency and the in-band spectral index can be well explained within 
the jet model. 

Similar to Sw J1644+57, the X-ray emission shows distinctive behavior
at late time compared with the expectation of the external shock 
synchrotron emission from the jet-CNM interaction, which suggests a
non-external-shock origin. Taking Sw J1666+57 as a template, we find that
the expected internal jet emission would outshine the observed flux greatly.
This suggests that the jet is off-beam. By requiring both the internal 
dissipation emission and the external shock X-ray fluxes not to exceed 
the observed values, we find that the initial Lorentz factor $\Gamma_j 
\gtrsim$ a few and $\theta_{\rm obs} \gtrsim 30^{\rm o}$. 
Our modeling therefore establishes IGR J12580+0134 as the first TDE with
an off-beam relativistic jet. The upcoming high resolution mapping from 
the Very Long Baseline Array (VLBA) will directly test this scenario and
image the jet structure.

It is also interesting to investigate a non-relativistic outflow model similar to that introduced by Alexander et al. (2015) for the radio emission in ASASSN-14li. We first model the radio data of IGR J12580+0134 with a non-relativistic spherical outflow in a uniform CNM. We find that the data require a kinetic energy of $\sim 10^{50} {\rm erg }$ and an initial velocity of $~0.3 c$. This model also needs a very high CNM density of $\sim 30 {\rm cm}^{-3}$ to allow for a significant deceleration of the outflow at $t \sim 300$ days. These parameters are not favored. We then consider an outflow model but with a circumnuclear gas density profile of $n \propto r^{-2}$. This model suggests a similar kinetic energy of $\sim 10^{50} {\rm erg} $, and an ejecta velocity of $~0.2 c$. However, due to the steep circumnuclear density profile, the outflow would undergo a coasting phase for years. As discussed in Section 2.2, the disrupted object is around $8 - 40$ Jupiter mass. If we take a typical mass $m_* \sim 0.01$, the outflow driven by the unbound tidal debris can then reach a velocity $v_{\rm ej} \simeq 0.014 c \ b m_{\bullet,6}^{1/6} m_{*,-2}^{19/48}$ and a kinetic energy $E_{\rm ej} \simeq 1.8 \times 10^{48} {\rm erg} b^2 m_{\bullet,6}^{1/6} m_{*,-2}^{43/24}$, both of which are much lower than the model requirements. Also such a strong outflow is unlikely launched from a sub-Eddington accretion disk. We therefore conclude that the off-beam relativistic jet model as proposed here is more natural to interpret the data of IGR J12580+0134.

In our modeling, the observed flux for an off-beam jet is reduced by a factor of $\sim a_{\rm off}^3$, assuming a point source (see Granot et al. 2002), which is a good approximation for a large view angle, $\theta_{\rm obs} \gg \theta_{rm j}$, as is our case.  For a near edge view, however, the correction on the flux would be $\sim a_{\rm off}$. This effect should be considered when studying the detection rate of TDEs (e.g. Sun et al. 2015).

We find it difficult to use the SSC counterpart of the radio emission from the jet to interpret the observed X-ray data. The rapid drop 
of the X-ray flux at $t>200 - 600$ days suggests that the X-ray emission 
likely comes from the internal dissipation of the jet or the accretion disk instead of the forward shock. An on-beam
geometry is also needed to account for the shape of the X-ray lightcurve 
within the SSC scenario, but such an on-beam geometry is disfavored by 
the lack of significant X-ray emission from the internal dissipation 
within the jet and the required very low Lorentz factor.

The observed X-ray emission at 40 d $\lesssim t \lesssim 300$ d exhibits 
the power-law decay with the index $\approx -5/3$, which is consistent with 
the predicted fallback rate of the standard TDE model. In this model,  
the disk luminosity is proportional to the accretion rate: $L_{\rm disk,X} 
\propto \dot{M}$. A natural source of the X-ray emission is the disk/corona 
around the BH, either the standard thin disk, or the advection-dominated 
accretion flow (ADAF) model. However, the ADAF model predicts $L_{\rm disk} 
\propto \dot{M}^2$. Therefore, the observations actually support the thin disk model. 
The existence of a corona near the thin disk also helps to facilitate 
the Blandford-Znajek mechanism (Blandford \& Znajek 1977; Lei et al. 2005; Lei et al. 2008), which is 
likely responsible for jet launching of other on-beam jetted
TDEs. 

Based on the current model, we predict that the radio flux should be $\sim 170 \ {\rm mJy}$ in L-band and $\sim 90 \ {\rm mJy}$ in C-band at present. If the CNM density has a stratified structure or if the electron index steepens with time, one would expect a somewhat lower flux then this predicted value.

\acknowledgments

We thank the anonymous referee for helpful suggestions and Dick Henriksen for useful comments.
This work is supported by National Basic Research Program (``973'' Program) 
of China under grant No. 2014CB845800, National Natural Science Foundation 
of China under grants U1431124, 11361140349 (China-Israel jointed program).

\clearpage


\newpage
\appendix
\section{Dynamical evolution and synchrotron emission in an off-beam relativistic jet}


To illustrate the main features of the dynamical evolution and radiation 
properties of an off-beam jet, we provide an analytical analysis in the 
following. The equations generally follow the recent review article of 
Gao et al. (2013) on the on-beam analytical synchrotron radiation models 
of GRBs. For an off-beam observer, the frequencies (times) should be 
multiplied (divided) by a factor of $a_{\rm off}$, and the fluxes should 
be corrected following Eq. (11). In the following equations, the 
dependence on $a_{\rm off}$ is explicitly presented. 
For the parameters given in Table 1, $a_{\rm off}$ is found to be around 
0.01 at early times and gradually increase to unity at late times. 
So it has a strong effect on the shape of the early-time lightcurve.

\subsection{Coasting Phase}
The relativistic jet first undergoes a coasting phase, in which we have 
$\Gamma (t) = \Gamma_\mathrm{j}$, and the distance of the shock front from the explosion center is $R(t) = 2c \Gamma_\mathrm{j}^2 t$. 
Based on the evolution of $\Gamma (t)$, we can give the expressions for 
the time evolution of the characteristic synchrotron frequencies (i.e., the minimum injection frequency $\nu_{\rm m}$, the cooling frequency $\nu_{\rm c}$ and the self-absorption frequency $\nu_{\rm a}$) as 
(Wu et al. 2003; Gao et al. 2013)
\begin{eqnarray}
&&\nu_{\rm m}=   3.6 \times 10^{12}~{\rm Hz}~a_{\rm off} n^{1/2} \Gamma_{\rm j,1}^{4} \epsilon_{e,-1}^{2}\epsilon_{\rm B,-1}^{1/2}           ,\nonumber\\
&&\nu_{\rm c}=    1.8 \times 10^{13}~{\rm Hz}~a_{\rm off}^{-1} n^{-3/2}\Gamma_{\rm j,1}^{-4}  \epsilon_{B,-1}^{-3/2} t_{d}^{-2}           ,\nonumber\\
&&\nu_{\rm a} =1.0 \times 10^{11}~{\rm Hz}~ a_{\rm off}^{8/5} n^{4/5} \Gamma_{\rm j,1}^{8/5} \epsilon_{e,-1}^{-1}\epsilon_{B,-1}^{1/5}t_{d}^{3/5},\nonumber\\
\label{eq:coasting}
\end{eqnarray}
in which the electron spectral index $p = 2.17$ is adopted. During this phase one has $\nu_a < \nu_m < 
\nu_c$. The radio emission undergoes a transition from being optically thin 
to optically thick at $\nu_a$. 

The synchrotron flux is given by
\begin{eqnarray}
&& F_\nu = 4.8\times 10^5 ~ {\rm mJy}~ a_{\rm off}^{17/3} n^{4/3} \Gamma_{\rm j,1}^{26/3} \epsilon_{\rm e,-1}^{-2/3}\epsilon_{\rm B,-1}^{1/3} \theta_{\rm j,-1}^2 \nu_{9}^{1/3} t_{\rm d}^3 , ~ \nu_a<\nu<\nu_m \nonumber \\ 
&& F_\nu = 205 ~ {\rm mJy}~ a_{\rm off}^3 \Gamma_{\rm j,1}^{6} \epsilon_{e,-1} \theta_{\rm j,-1}^2 \nu_{9}^{2} t_{\rm d}^2
, ~~~~~~~~~~~~~~~~~~~~~~~~~~ \nu<\nu_a<\nu_m. \nonumber \\
\end{eqnarray}

\subsection{Deceleration Phase}

The jet starts to decelerate when the mass of the CNM swept by the forward
shock is about $1/\Gamma_{\rm j}$ of the rest mass in the ejecta. The 
deceleration time of the ejecta with an isotropic kinetic energy 
$E_{\rm k,iso}$ and an initial Lorentz factor $\Gamma_{\rm j}$ is
\begin{equation}
t_{\rm dec} = a_{\rm off}^{-1} (1+z) \left[ \frac{3E_{\rm k,iso} }{16 \pi n m_{\rm p} \Gamma_{\rm j}^8 c^5 } \right]^{1/3} \simeq   0.13 \ {\rm day} \ a_{\rm off}^{-1} n^{-1/3} E_{50}^{1/3}  \Gamma_{\rm j, 1}^{-8/3},
\end{equation}
where $E_{50}$ denotes $E_{\rm k,iso}$ in units of $10^{50}$ erg. The off-beam
correction factor ($a_{\rm off}$) makes the deceleration time longer in the observer frame.
After $t_{\rm dec}$, the jet approaches the Blandford 
\& McKee (1976) self-similar evolution, with
\begin{eqnarray}
&& \Gamma (t) \simeq 2.1  a_{\rm off}^{-3/8} E_{ 50}^{1/8}  t_{\rm d}^{-3/8}, \nonumber\\
&& R (t) \simeq 1.2 \times 10^{17}  \ {\rm cm} \  a_{\rm off}^{1/4}  E_{ 50}^{1/4}  t_{\rm d}^{1/4}.
\label{BM}
\end{eqnarray}
At this stage, the characteristic synchrotron frequencies are give by 
(Gao et al. 2013)
\begin{eqnarray}
&&\nu_{\rm m}=   6.5\times 10^{9}~{\rm Hz}~ a_{\rm off}^{-1/2} E_{50}^{1/2}\epsilon_{\rm e,-1}^{2}\epsilon_{\rm B,-1}^{1/2} t_{\rm d}^{-3/2}           ,\nonumber\\
&&\nu_{\rm c}=    1.0\times 10^{16}~{\rm Hz}~ a_{\rm off}^{1/2} n^{-1} E_{50}^{-1/2} \epsilon_{B,-1}^{-3/2}t_{\rm d}^{-1/2},           \nonumber\\
&&\nu_{\rm a}= 2.9\times 10^{10}~{\rm Hz}~ a_{\rm off} n^{3/5} E_{50}^{1/5} \epsilon_{\rm e,-1}^{-1} \epsilon_{\rm B,-1}^{1/5}, ~~~~~~~~~~
\nu_{\rm a} < \nu_{\rm m} < \nu_{\rm c} \nonumber\\
&&\nu_{\rm a}= 1.5 \times 10^{10}~{\rm Hz}~ a_{\rm off}^{0.31} n^{0.32} E_{50}^{0.34} \epsilon_{\rm e,-1}^
{0.38} \epsilon_{\rm B,-1}^{0.34} t_{\rm d}^{-0.69}, ~\nu_{\rm m} < \nu_{\rm a} < \nu_{\rm c} . \nonumber \\
\end{eqnarray}
One can see that $\nu_m$ decreases very quickly with time. Therefore the 
jet would evolve from the regime $\nu_a<\nu_m<\nu_c$ to $\nu_m<\nu_a<\nu_c$ 
after the deceleration.

During the first several days, the jet decelerates in the $\nu_a<\nu_m<\nu_c$ 
regime. We have the synchrotron flux
\begin{eqnarray}
&&F_\nu = 1.0  ~{\rm mJy}~ a_{\rm off}^{3/4} n^{-0.75} E_{50}^{0.75} \epsilon_{\rm e,-1}  \theta_{\rm j,-1}^2 \nu_{9}^{2} t_{\rm d}^{-1/4}, ~~~~~~~~~~ \nu<\nu_{\rm a}<\nu_{\rm m}<\nu_{\rm c} \nonumber \\
&&F_\nu = 271  ~{\rm mJy}~ a_{\rm off}^{29/12} n^{1/4} E_{50}^{13/12} \epsilon_{\rm e,-1}^
{-2/3} \epsilon_{\rm B,-1}^{1/3}  \theta_{\rm j,-1}^2 \nu_{9}^{1/3} t_{\rm d}^{-1/4}, ~ \nu_{\rm a}<\nu<\nu_{\rm m}<\nu_{\rm c}. \nonumber \\
\end{eqnarray}
It then evolves to the $\nu_m<\nu_a<\nu_c$ regime in about ten days. 
The flux is
\begin{eqnarray}
&&F_\nu = 1.0  ~{\rm mJy}~ a_{\rm off}^{3/4} n^{-0.75} E_{50}^{0.75} \epsilon_{\rm e,-1}  \theta_{\rm j,-1}^2 \nu_{9}^{2} t_{\rm d}^{-1/4}, ~~~~~~~~~~ \nu<\nu_m<\nu_a<\nu_c \nonumber \\
&&F_\nu = 0.4  ~{\rm mJy}~ a_{\rm off} n^{-0.75} E_{50}^{0.5}  \epsilon_{\rm B,-1}^{-1/4}  \theta_{\rm j,-1}^2 \nu_{9}^{5/2} t_{\rm d}^{0.5}, ~~~~~~~~~~ \nu_m<\nu<\nu_a<\nu_c \nonumber \\
&&F_\nu = 1517 ~{\rm mJy}~ a_{\rm off}^2 n^{0.25} E_{50}^{1.54} \epsilon_{\rm e,-1}^
{1.17} \epsilon_{\rm B,-1}^{0.79} \theta_{\rm j,-1}^2 \nu_{9}^{-0.585} t_{\rm d}^{-1.63}, ~ \nu_{\rm a}<\nu<\nu_{\rm c}.  \nonumber \\
\end{eqnarray}

\subsection{Newtonian Phase}

The blastwave eventually enters the Newtonian phase when it has swept 
up the CNM with the total rest mass energy comparable to the energy of 
the ejecta. This Sedov time is
\begin{equation}
t_{\rm Sedov} = (1+z) \frac{3}{17} \left[ \frac{3E_{\rm k,iso} }{4 \pi n m_{\rm p}  c^5 } \right]^{1/3} \simeq 17 \ {\rm day} \   n^{-1/3} E_{50}^{1/3}.
\end{equation}
In the non-relativistic (Newtonian) regime, the dynamics is described 
by the well know Sedov-Taylor solution. The factor $a_{\rm off} \simeq 1$ in this stage. We have the synchrotron frequencies 
as (Gao et al. 2013)
\begin{eqnarray}
&&\nu_m=   3.8\times10^{12}~{\rm Hz}~n^{-1/2} E_{50} \epsilon_{\rm e,-1}^{2}\epsilon_{\rm B,-1}^{1/2}t_{\rm d}^{-3}           ,\nonumber\\
&&\nu_c=    3.7\times10^{15}~{\rm Hz}~n^{-0.9} E_{50}^{-3/5}\epsilon_{\rm B,-1}^{-3/2}t_{\rm d}^{-1/5} ,          \nonumber\\
&&\nu_a= 2.7 \times10^{10}~{\rm Hz}~n^{0.31} E_{50}^{0.35} \epsilon_{\rm e,-1}^{0.38}\epsilon_{\rm B,-1}^{0.34} t_{\rm d}^{-0.73},
\label{eq:Newton_nu}
\end{eqnarray}
with $\nu_{\rm m} < \nu_{\rm a} < \nu_c$. The synchrotron flux in this regime
can be written as
\begin{eqnarray}
&& F_\nu = 720 ~{\rm mJy}~ n^{0.41} E_{50}^{1.385} \epsilon_{\rm e,-1}^{1.17}  \epsilon_{\rm B,-1}^{0.79} \theta_{\rm j,-1}^2 \nu_{9}^{-0.585} t_{\rm d}^{-1.16} , ~ \nu_a<\nu<\nu_c  \nonumber \\
&& F_\nu = 0.029 ~{\rm mJy}~ n^{-0.55} E_{50}^{0.3}  \epsilon_{\rm B,-1}^{-0.25} \theta_{\rm j,-1}^2 \nu_{9}^{5/2} t_{\rm d}^{1.1}, ~~~~~~~~~~~ \nu_{\rm m}<\nu<\nu_{\rm a}.
\end{eqnarray}



\begin{thebibliography}{}
\bibitem[Alexander et al.(2015)]{abg15} Alexander, K. D., Berger, E., Guillochon, J., Zauderer, B. A. \& Williams, P. K. G. 2015, arXiv:1510.01226

\bibitem[Ayal et al.(2000)]{alp00} Ayal, S., Livio, M., \& Piran, T. 2000, ApJ, 545, L143

\bibitem[{Bardeen et al.}(1972)]{Bardeen72} Bardeen J. M., Press W. H., \& Teukolsky S. A. 1972, ApJ, 178, 347

\bibitem[Beckert \& Falcke(2002)]{bf02} Beckert, T. \& Falcke, H. 2002, A\&A, 388, 1106

\bibitem[Bian \& Huang(2010)]{bh10} Bian, W. \& Huang, K. 2010, MNRAS, 401, 507

\bibitem[Blandford \& McKee(1976)]{bm76} Blandford, R. D., \& McKee, C. F. 1976, Phys. Fluids, 19, 1130

\bibitem[Blandford \& Znajek(1977)]{bz77} Blandford, R. D., \& Znajek, R. L. 1977, MNRAS, 179, 433

\bibitem[Bloom et al.(2011)]{bgm11} Bloom, J. S., Giannios, D., Metzger, B. D., et al. 2011, Science, 333, 203

\bibitem[Brown et al.(2015)]{bls15} Brown, G. C., Levan, A. J., Stanway, E. R., et al. 2015, arXiv:1507.03582

\bibitem[Burrows et al.(2011)]{bkg11} Burrows, D. N., Kennea, J. A., Ghisellini, G., et al. 2011, Nature, 476, 421

\bibitem[Cenko et al.(2012)]{ckh12} Cenko, S. B., Krimm, H. A., Horesh, A., et al. 2012, ApJ, 753, 77

\bibitem[Chabrier \& Baraffe(2000)]{cb00} Chabrier, G., \& Baraffe, I. 2000, ARA\&A, 38, 337

\bibitem[Evans \& Kochanek(1989)]{ek89} Evans, C. R., \& Kochanek, C. 1989, ApJ, 346, L13

\bibitem[Fabbiano et al.(1992)]{fkt92} Fabbiano, G., Kim, D.-W., \& Trinchieri, G. 1992, ApJS, 80, 531

\bibitem[Gao et al.(2013)]{glzwz13} Gao, H., Lei, W. H., Zou, Y. C., Wu, X. F., \& Zhang, B. 2013, New Astron. Rev. , 57, 141

\bibitem[Granot et al.(2002)]{gpkw02} Granot, J., Panaitescu, A, Kumar, P., \& Woosley, S. E. 2002, ApJL, 570, 61

\bibitem[Greene \& Ho(2005)]{gh05} Greene, J. E. \& Ho, L. C. 2005, ApJ, 630, 122

\bibitem[Haring \& Rix(2004)]{hr04} H{\"a}ring, N. \& Rix, H. W. 2004, ApJL, 604, L89

\bibitem[Huang et al.(2000)]{hgdl00} Huang, Y. F., Gou, L. J., Dai, Z. G., \& Lu, T. 2000, ApJ, 543, 90

\bibitem[Huang \& Cheng(2003)]{hc03} Huang, Y. F. \& Cheng, K. S. 2003, MNRAS, 341, 263

\bibitem[Ho et al.(1997)]{hfs97} Ho, L. C., Filippenko, A. V., \& Sargent, W. L. W. 1997, ApJS, 112, 315

\bibitem[Irwin et al.(2015)]{Irwin15} Irwin, J. A., Henriksen, R. N., Krause, M., Wang, Q. D., Wiegert, T., Murphy, E. J., Heald, G., \& Perlman, E. 2015, ApJ, 809, 172

\bibitem[Kumar \& Zhang(2015)]{kz15} Kumar, P. \& Zhang, B. 2015, Phys. Rep., 561, 1

\bibitem[Kormendy \& Gebhardt(2001)]{kg01} Kormendy, J. \& Gebhardt, K. 2001, AIPC, 586, 363

\bibitem[Lei et al.(2005)]{Lei05} Lei, W. H., Wang, D. X. \& Ma, R. Y. 2005, ApJ, 619, 420

\bibitem[Lei et al.(2008)]{Lei08} Lei, W. H., Wang, D. X., Zou, Y. C. \& Zhang, L. 2008, ChJAA, 8, 404

\bibitem[Lei \& Zhang(2011)]{lz11} Lei, W. H., \& Zhang, B. 2011, ApJ, 740, L27

\bibitem[Lei et al.(2013)]{lzg13} Lei, W. H., Zhang, B., \& Gao, H. 2013, ApJ, 762, 98

\bibitem[Levan et al. (2011)]{ltc11} Levan, A. J., Tanvir, N. R., Cenko, S. B., et al. 2011, Science, 333, 199

\bibitem[Li et al.(2002)]{lnm02} Li, L. X., Narayan, R., \& Menou, K. 2002, ApJ, 576, 753 

\bibitem[Liu et al. (2015)]{lpl15} Liu, D. B., Peer, A., \& Loeb, A. 2015, ApJ, 798, 13

\bibitem[Magorrian \& Tremaine(1999)]{mt99} Magorrian, J. \& Tremaine, S. 1999, MNRAS, 309, 447

\bibitem[Metzger et al. (2012)]{mgm12} Metzger, B. D., Giannios, D., \& Mimica, P. 2012, MNRAS, 420, 3528

\bibitem[M{\"i}ller \& G{\"u}ltekin(2011)]{mg11} M{\"i}ller, J. M. \& G{\"u}ltekin, K. 2011, ApJL, 738, L13

\bibitem[Nikolajuk \& Walter(2013)]{nw13} Nikolajuk, M., \& Walter, R. 2013, A\&A, 552, A75 

\bibitem[Novikov \& Thorne(1973)]{nt73} Novikov, I. D., \& Thorne, K. S. 1973, in Black Holes, ed. C. DeWitt-Morette \& B.S.DeWitt(NewYork:Gordon \& Breach),345

\bibitem[O'Sullivan et al. (2013)]{omf13} O'Sullivan, S. P., McClure-Griffiths, N. M., Feain, I. J. et al. 2013, MNRAS, 435, 311

\bibitem[Phinney(1989)]{Phinney89} Phinney, E. 1989, IAU Symposium, 136, 543

\bibitem[Rees(1988)]{Rees88} Rees, M. J. 1988, Nature, 333, 523

\bibitem[Remillard \& McClintock(2006)]{rm06} Remillard R. A. \& McClintock J., E. 2006, ARA\&A, 44, 49

\bibitem[Shen \& Matzner(2014)]{sm14} Shen, R. F., \& Matzner, C. D. 2014, ApJ, 784, 87

\bibitem[Sironi \& Giannios(2013)]{sg13} Sironi, L. \& Giannios, D. 2013, ApJ, 778, 107

\bibitem[Solanes et al.(2002)]{Solanes02} Solanes, J. M., Sanchis, T., Eduard Salvador-Sole, E., et al. 2002, AJ, 124, 2440

\bibitem[Sun, Zhang \& Li(2015)]{szl15} Sun, H., Zhang, B. \& Li, Z. 2015, ApJ, 812, 33

\bibitem[Tchekhovskov et al. (2014)]{tmgk14} Tchekhovskoy, A., Metzger, B. D., Giannios, D., \& Kelley, L. Z. 2014, MNRAS, 437, 2744

\bibitem[Walter et al.(2011)]{Walter11} Walter, R. et al. 2011, ATel, 3108

\bibitem[Wang et al.(1998)]{wly98} Wang, D. X., Lu, Y., Yang, L. T., 1998, MNRAS, 294, 667

\bibitem[Wang \& Merritt(2004)]{wm04} Wang, J. \& Merritt, D. 2004, ApJ, 600, 149

\bibitem[Wang et al.(2014)]{Wang14} Wang, J. Z., Lei, W. H., Wang, D. X., Zou, Y. C., Zhang, B., Gao, H., \& Huang, C. Y. 2014, ApJ, 788, 32

\bibitem[Wu et al. (2011)]{wzc11} Wu, Q., Zou, Y. C., Cao, X., Wang, D. X., Chen, L. 2011, ApJ,
740, 21

\bibitem[Wu et al. (2015)]{wzl15} Wu, Q., Zhang, B., Lei, W. H., Zou, Y. C., Liang, E. W., Cao, X. 2015, arXiv: 1509.04896

\bibitem[Zauderer et al.(2011)]{zbs11} Zauderer, B. A., Berger, E., Soderberg, A. M., et al. 2011, Nature, 476, 425

\bibitem[Zauderer et al.(2013)]{zbm13} Zauderer, B. A., Berger, E., Margutti, R., et al. 2013, ApJ, 767, 152

\bibitem[Zhang \& M\'esz\'aros(2001)]{zm01}Zhang, B., \& M\'esz\'aros, P. 2001, ApJ, 552, L35

\bibitem[Zhang \& M\'esz\'aros(2004)]{zm04}Zhang, B., \& M\'esz\'aros, P. 2004, International Journal of Modern Physics A, 19, 2385

\bibitem[Zhang \& Yan(2011)]{zy11} Zhang, B., \& Yan, H. 2011, ApJ, 726, 90


\end{thebibliography}
\end{document}